\documentclass[aps,prd,twocolumn,superscriptaddress,showpacs,nofootinbib]{revtex4}
\usepackage{natbib,hyperref,ifthen}
\usepackage{graphicx}
\usepackage{dcolumn}
\usepackage{bm}
\usepackage{natbib}
\usepackage{multirow}
\usepackage{epsfig}
\usepackage{amsmath}

\newcommand{\beq}{\begin{equation}}
\newcommand{\eeq}{\end{equation}}
\newcommand{\beqa}{\begin{eqnarray}}
\newcommand{\eeqa}{\end{eqnarray}}

\newcommand{\be}{\begin{equation}}
\newcommand{\ee}{\end{equation}}

\newcommand{\rmd}{{d}}
\renewcommand{\vec}[1]{{\bf #1}}

\newcommand{\fnl}{{f_{\rm NL}}}

\def\be{\begin{equation}}
\def\ee{\end{equation}}
\newcommand{\bea}{\begin{eqnarray}}
\newcommand{\eea}{\end{eqnarray}}

\def\fNL{f_{\rm NL}}
\def\fnl{f_{\rm NL}}

\def\sh{E'-B}
\def\H{{\cal H}}
\def\A{A}

\begin{document}

\title{Scale-dependent bias from primordial non-Gaussianity in general relativity}

\author{David Wands} \affiliation{Institute of Cosmology and
Gravitation, University of Portsmouth, Dennis Sciama Building,
Portsmouth, PO1 3FX, United Kingdom}

\author{An\v{z}e Slosar} \affiliation{Berkeley Center for Cosmological
   Physics, Physics Department and Lawrence Berkeley National
   Laboratory,University of California, Berkeley California 94720, USA}
\affiliation{Faculty of Mathematics and Physics, University of
   Ljubljana, Slovenia}

\date{\today}

\begin{abstract}
In this note we examine the derivation of scale-dependent bias due
to primordial non-Gaussianity of the local type in the context of
general relativity.  We justify the use of the Poisson equation in
general relativistic perturbation theory and thus the derivation of
scale-dependent bias as a test of primordial non-Gaussianity, using
the spherical collapse model. The corollary is that the form of
scale-dependent bias does not receive general relativistic
corrections on scales larger than the Hubble radius. This leads to a
formally divergent correlation function for biased tracers of the
mass distribution which we discuss.
\end{abstract}


\maketitle

\setcounter{footnote}{0}

\section{Introduction}

One of the most powerful discriminants between various models for
the origin of structure in the Universe is the non-Gaussianity of
the primordial fluctuations. The non-linear evolution of
inhomogeneities on super-Hubble scales during or after slow-roll
inflation can give rise to the local form of non-Gaussianity where
the primordial curvature perturbation can be given as a local
function of a Gaussian random field
\citep{1990PhRvD..42.3936S,1994ApJ...430..447G,2000MNRAS.313..141V,2005PhRvL..95l1302L}.
This occurs, for example in the curvaton scenario, due to the
late-decay of a weakly coupled scalar field after inflation
\cite{2003PhRvD..67b3503L}, or in the new ekpyrotic models of an
accelerated collapse with two or more fields
\cite{2007JCAP...11..010C,2007JCAP...11..024K,2008PhRvL.100q1302B,2008PhRvD..77f3533L}.

Local non-Gaussianity is commonly characterised by a single
parameter, $\fNL$, where the local primordial Newtonian potential on
large scales in the matter-dominated era is given by
\citep{2001PhRvD..63f3002K}
\begin{equation}
  \label{phiNfNL}
  \phi_N(x) = \phi_G(x) + \fNL \left( \phi_G^2(x) - \langle \phi_G^2 \rangle \right) \,,
\end{equation}
and $\phi_G(x)$ is a Gaussian random field\footnote{Note that we
adopt
   the sign convention of Komatsu \& Spergel
   \citep{2001PhRvD..63f3002K} which is widely used to characterise
   primordial non-Gaussianity; the Newtonian potential is positive for
   a point mass $\phi_N(r)=+G m/r$. Many theoretical calculations of
   primordial non-Gaussianity, starting with that of Maldacena
   \citep{2003JHEP...05..013M}, use the opposite sign for the Newtonian
   potential, as for instance used in the influential review by
   Mukhanov, Feldman and Brandenberger \cite{1992PhR...215..203M}, and
   this leads to a different sign for $\fNL$. Although the choice of
   sign is conventional, observations are sensitive to the sign of
   $\fNL$.}.

In Newtonian gravity the Poisson equation,
 \be
 \label{firstPoisson}
   - \nabla^2 \phi_N = 4\pi G \rho \delta\,,
 \ee
gives the local density contrast, $\delta$, as a function not of the
local Newtonian potential directly, but rather as a function of the
spatial divergence of the Newtonian potential. A consequence of
combining equation (\ref{phiNfNL}) with the Poisson equation
(\ref{firstPoisson}) is that the non-Gaussian part of the density
field is not a local function of the Gaussian part.

If we re-write the physical Laplacian in terms of comoving
coordinates in a Friedmann-Robertson-Walker (FRW) universe,
$\nabla^2=\partial^2/a^2$, where $a$ is the scale factor, we can
write the Poission equation as \be
  \label{Poisson}
- \partial^2 \phi_N = 4\pi G a^2 \rho \delta\,. \ee The Fourier
transform of this equation gives
\begin{equation}
   k^2 \phi_N(\vec{k}) = \frac{3}{2}a^2H^2 \delta({\vec{k}}) \,
\end{equation}
where we use the Friedmann equation to give the background density
in terms of the Hubble expansion, $H$. It is the scale-dependence of
this relation which is ultimately responsible for the diverging
scale-dependent bias due to non-Gaussianity as $k\to0$.

It is commonly assumed that the formation of astrophysical objects
is a local function of the dark matter overdensity (smoothed over
some scale $R$):
\begin{equation}
   \delta_g(\vec{x}) = f(\delta (\vec{x}), (\nabla \delta (\vec{x}))^2,
     \nabla^2 \delta (\vec{x}), \ldots) + s(\vec{x}),
\end{equation}
where $s(\vec{x})$ is a non-deterministic stochastic component. Iff
$R$ is large enough so that $\delta$ is small and if Fourier modes are
decoupled, then the overdensity of astrophysical objects traces dark
matter fluctuations in the Fourier space in the limit of $k\rightarrow
0$:
\begin{equation}
   \delta_g(\vec{k}) = b \delta (\vec{k}) + s(\vec{k}),
\end{equation}
where $b=f'(0)$ is the so-called bias parameter.  The corresponding
power spectrum is given by
\begin{equation}
   P_{gg}(k) = b^2 P(k) + \sigma_s^2.
\end{equation}
The stochastic white noise tail has never been observed and hence we
will neglect it here. The important point is that bias $b$ should
approach a constant as $k$ approaches zero. Deviation from this
implies either that the formation process is non-local or that
Fourier modes are coupled, i.e.,
the dark matter density fluctuations non-Gaussian.

In concrete models of formation of astrophysical objects, it is
assumed that the function $f$ is such, that it puts objects in the
peaks of the local density field \cite{1974ApJ...187..425P}. Such
density peaks collapse under their own gravity and form virialized
objects. The local number density of such objects in Lagrangian
space is given by
\begin{equation}
   n = n_0 (1 + b_L \delta_l),
\end{equation}
where $b_L$ stands for Lagrangian space bias, $\delta_l$ is the
contribution from the very long wavelength modes that essentially
modulate the mean density of the effective local cosmology. Therefore
\begin{equation}
   b_L = n_0^{-1}\frac{\partial n}{\partial \delta_l}
\end{equation}
and the more usual Eulerian-space bias is given by $b=1+b_L$
\cite{1988MNRAS.235..715E,1989MNRAS.237.1127C,1996MNRAS.282..347M,1998MNRAS.297..692C}.
This approach towards formation of astrophysical objects is the
essence of the so-called peak-background split. If one understands
how the local effective cosmology is modulated by the presence of
the large-scale modes and how the formation of objects is affected
by the change in the local effective cosmology, then one can
calculate the bias parameter for that particular tracer.

The bias is therefore given by the interplay between the small-scale
collapse dynamics and the large scales, which modulate the local
cosmology through fluctuations present in the very large scale
modes. The purpose of this paper is to analyse how can this be viewed
within the context of general relativity.


The distinctive scale-dependence of large-scale structure has
recently been proposed as a powerful test of the local form of
primordial non-Gaussianity \citep{Dalal:2007cu}.
For a biased tracer of structure, it can be shown that local
non-Gaussianity induces a scale dependent bias $b=b_G+\Delta b$,
with \citep{2008JCAP...08..031S}:
\begin{equation}
   \Delta b = \frac{2 \fnl}{\alpha(k)} \frac{\partial \log n}{\partial
     \log \sigma_8},
\end{equation}
where
  \be
\alpha(k) = \frac{2 c^2 T(k) D(z)}{3 \Omega_m H_0^2}k^2 \,,
 \ee
is the Fourier space conversion factor between the present day
density perturbations and the Newtonian potential during matter
domination on large scales, including the linear transfer function
$T(k)$, which is approaches unity on large scales, and growth factor
normalised to be $D(z)= 1/(1+z)$ in the matter domination. $\partial
\log n/\partial \log \sigma_8$ describes how the number density of
astrophysical objects, $n$, is affected by the change of the
amplitude of small-scale fluctuations $\sigma_8$ \emph{in Gaussian
cosmologies}. Therefore, although the initial derivation relied on
the statistics of Gaussian fields, it is now clear that the effect
is present for all local theories for the formation of the
astrophysical objects \citep{2008ApJ...677L..77M,
2008arXiv0805.3580S,
   2008arXiv0806.1046A,2008arXiv0806.1061M}. In particular, for a
universal form of the mass function $n(M)$, the above equation
simplifies to
\begin{equation}
   \Delta b = \fnl (b_G-1) \frac{3 \delta_* \Omega_m H_0^2}{c^2 k^2 T(k) D(z)},
\label{eq:4}
\end{equation}
where $b_G$ is the Gaussian bias and $\delta_* \sim 1.68$ is the
linear over-density at collapse for the spherical collapse model and
other symbols have their usual meaning.

A striking feature of the Equation (\ref{eq:4}) is that the
scale-dependent correction diverges as the wave-vector $k$
approaches zero. This is a direct consequence of the Newtonian form
of the Poisson equation in the Fourier space.  Since the Newtonian
approximation is expected to break down at the scales comparable to
or larger than the Hubble scale, it is timely to ask whether the
divergence of the scale-dependent bias is real or an artefact of the
Newtonian approximation. This is the main question that we address
in this brief report.

We start by analysing the Poisson equation in the context of general
relativistic perturbation theory in Section
\ref{sec:gr-poisson-equation}. Next we review the well known
\emph{exact} general relativistic solution, namely the spherical
collapse solution (Section \ref{sec:spherical-collapse}).  By
investigating the metric in the limit of small perturbations, we can
identify the gauge in which the spherhical collapse calculation is
performed.  This justifies the form the Poisson equation for linear
perturbations on all scales.  In Section
\ref{sec:correlation-function} we discuss the formally divergent
correlation function of a biased tracer of structure and show how
this is absent in observations of real correlation functions. We
conclude in Section \ref{sec:conclusions}.


\section{General relativistic Poisson equation}
\label{sec:gr-poisson-equation}

One might naively expect corrections to the Newtonian Poisson
equation (\ref{Poisson}) of order $(aH/k)^2$ which would become
large close to the Hubble scale and change the scale-dependence of
$\alpha(k)$ on large scales. However, the form of the general
relativistic constraint equations is gauge dependent.

In a completely homogeneous universe, the FRW metric $ds^2 = dt^2 -
a^2 \delta_{ij} dx^{i} dx^j= a^2 \left(d\eta^2 -\delta_{ij} dx^{i}
   dx^j\right)$ has no gauge freedom. At the first order of
perturbations, on the other hand, the coordinate freedom gives rise
to ambiguities in the definition of the density perturbation in an
inhomogeneous spacetime (see, for instance,
Ref.~\cite{2008arXiv0809.4944M}). In particular the freedom to make
an inhomogeneous redefinition of the time coordinate $\eta\to
\eta+\delta\eta(x,\eta)$ leads to a redefinition of the density
perturbation $\delta\rho\to\delta\rho-\rho'\delta\eta$, where a
prime denotes derivatives with respect to conformal time, $\eta$.

The most general scalar perturbation\footnote{We neglect vector and
   tensor perturbations which are decoupled from scalar perturbations
   at first order.}  of the spatially flat Friedman-Robertson-Walker
(FRW) metric can be written as
\cite{1992PhR...215..203M,2008arXiv0809.4944M}
\begin{multline}
  \label{pertFRW}
ds^2 = a^2 \bigl\{ (1+2\A) d\eta^2 - 2(\partial_iB)dx^id\eta \\
   - [ (1-2\psi)\delta_{ij} + 2(\partial_i\partial_jE) ] dx^idx^j \bigr\} \,.
\end{multline}
but the metric perturbations $\A$, $B$, $\psi$ and $E$ are all, like
the density perturbation, $\delta\rho$, gauge-dependent.

In an arbitrary gauge there is a first-order energy constraint
\be \label{eq:densitycon}
  - 3\H\left(\psi'+\H\A\right) + \partial^2\left[\psi+\H(\sh)\right] = 4\pi G a^2 \delta\rho \,.
\ee where $\H=aH$ is the conformal Hubble rate, and a first-order
momentum constraint
  \be
  \partial_i \left( \psi'+\H\A \right) = -4\pi G a^2 \partial_i (\delta q)
  \label{eq:mtmcon}
\ee
where $\partial_i(\delta q)$ is the 3-momentum. We can combine these
two equations to obtain a general relativistic Poisson equation \be
  \label{GRPoisson}
  \partial^2\left[\psi+\H(\sh)\right] = 4\pi G a^2 \left[ \delta\rho
    -3\H \delta q \right] \,.
\ee

This is exactly the same as the Newtonian Poisson equation
(\ref{Poisson}) as long as we identify $-\phi_N$ with the metric
potential in the Newtonian or longitudinal gauge \be
  \label{phiN}
- \phi_N = \psi_\ell \equiv \psi + \H(\sh) \,, \ee and $\rho\delta$
with the density perturbation in the comoving-orthogonal gauge
  \be
  \label{deltarhoc}
  \rho \delta = \delta\rho_c \equiv \delta\rho - 3\H \delta q \,.
  \ee
i.e., the density perturbation in a gauge in which $B=0$ and $\delta
q=0$.

The general relativistic Poisson equation (\ref{GRPoisson}) relates
gauge-invariant combinations of metric and matter perturbations,
which nonetheless have the interpretation as potential or density
perturbations in two different gauges
\citep{1980PhRvD..22.1882B,1998PhRvD..57.3290H}.

If the primordial Newtonian potential, $\phi_N$, has a local
non-Gaussianity as given in Equation~(\ref{phiNfNL}) then the
comoving-orthogonal density contrast, $\delta$, will indeed have a
non-local non-Gaussianity.

Note however, that in some gauges there are general relativistic
corrections to the Poisson equation. For example, in the
longitudinal gauge the shear potential, $\sh$, vanishes and on large
scales the regular solution, $-\phi_N=\A_\ell = \psi_\ell = {\rm
const}$, gives a relativistic correction to the Poisson equation
  \be
  \label{longPoisson}
3 \H^2 \phi_N - \partial^2 \phi_N = 4\pi G a^2 \delta\rho_\ell \,.
  \ee
Thus on super-Hubble scales ($|\partial^2\phi_N|\ll \H^2|\phi_N|$)
the density contrast in the longitudinal gauge,
$\delta_\ell=\delta\rho_\ell/\rho$, is proportional to the Newtonian
potential
  \begin{multline}
   \delta_\ell(x) \simeq - 2\phi_N (x)\\
    = -2 \phi_G(x) - 2\fNL \left( \phi_G^2(x) - \langle \phi_G^2 \rangle \right)
  \,.
  \end{multline}
This density contrast, $\delta_\ell$, {\em does} have a local
non-Gaussianity in the large scale limit.

Both the density perturbation in the comoving-orthogonal and
longitudinal gauge can be given as gauge-invariant combinations of
gauge-dependent variables, and on sub-Hubble scales
($|\partial^2\phi_N|\gg \H^2|\phi_N|$) we find from
Eqs.~(\ref{longPoisson}) and~(\ref{GRPoisson}) that
$\delta\rho_\ell\simeq \delta\rho_c$ and these two possible
definitions of the density perturbation coincide. The question then
arises as to which is the relevant density perturbation from which
to calculate the bias of galaxies and large-scale structure.

\section{Spherical collapse}
\label{sec:spherical-collapse}

The usual Newtonian calculation of the abundance of collapsed
objects such as galaxies or quasars assumes that the collapse is a
function of the local density field smoothed on some scale $R$. But
which is the appropriate density variable as we approach the Hubble
scale? Although initial conditions are set at early times where the
linear perturbation theory is valid, collapse requires a non-linear
calculation which is generally only possible in general relativity
if we make some assumption of simplifying symmetry. The most simple
example is that of spherical collapse. In the absence of any
pressure perturbation (as for pressureless matter, with or without a
cosmological constant) the model allows us to compute the non-linear
collapse of a top-hat overdensity described by a closed FRW model
embedded within a flat FRW exterior.

The Friedman equation for the unperturbed outer universe is given
by:
  \begin{equation}
  \label{Friedmann}
    H^2=\left(\frac{\dot{a}}{a}\right)^2 = \frac{8\pi G \rho_0}{3a^3},
  \end{equation}
while the inner region has a perturbed scale factor, $\tilde{a}$,
and small positive curvature, $K$, and hence
  \begin{equation}
  \label{pertFriedmann}
     \tilde{H}^2=\left(\frac{\dot{\tilde{a}}}{\tilde{a}}\right)^2 =
     \frac{8\pi G \rho_0}{3\tilde{a}^3} -\frac{K}{\tilde{a}^2}.
  \end{equation}

Without the loss of generality, we can set $8 \pi G \rho_0/3=1$.
The Friedman equation for the closed inner universe can then be
expressed in a parametric form
  \begin{eqnarray}
    \tilde{a} = \frac{1}{2 K} (1- \cos \theta)\\
    t = \frac{1}{2 K^{3/2}}(\theta-\sin \theta).
\end{eqnarray}
Expanding $\tilde{a}$ and ${t}$ to fifth order in $\theta$, one can
show that in the limit of $\theta \ll 1$,
  \begin{equation}
    \tilde{a} = \left(\frac{3t}{2} \right)^{2/3}\left(1-\frac{1}{20}\theta^2\right)
  \end{equation}
We see that to leading order the scale factors match as $\tilde{a}$
  approaches zero, if we have $a=(3t/2)^{2/3}$ as the
  solution for the exterior flat universe.

The linear over-densities are thus given by
  \begin{equation}
   \label{lineardelta}
    \delta = \frac{a^3}{\tilde{a}^3} -1 \simeq -3\frac{\delta a}{a} \simeq \frac{3}{20} \theta^2 \simeq \frac{3}{5} K a,
  \end{equation}
where $\delta a = \tilde{a}-a$. We have thus recovered the expected
result that linear perturbations grow with $a$ in the matter era and
that this spherically symmetrical system has a one parameter
solution.
%
%
When the inner perturbation collapses, $\theta_*=2\pi$,
$t_*=\pi K^{-3/2}$
and the linear density at that moment is
\begin{equation}
 \label{deltac}
    \delta_* = \frac{3}{5}\left(\frac{3\pi}{2}\right)^{2/3}\sim 1.68,
\end{equation}
which is the standard result.

But to decide which linear density perturbation this is, we need to
identify the coordinate choice implicit in the spherical collapse
model.
Let us look at the same process from the metric perturbation
perspective. Inside the overdense region the perturbed, non-linear
metric is given by a closed FRW metric
  \be
   ds^2 = dt^2 - \tilde{a}^2 \left[
   \frac{dr^2}{1-Kr^2} + r^2 d\Omega^2 \right] \,,
    \ee
where $K>0$.
For $Kr^2 \ll 1$, this can be recast in Cartesian form
  \be
  \label{pertds2}
   ds^2 = dt^2 - \tilde{a}^2 \left[ \delta_{ij} + Kr^2(\partial_i r)(\partial_j r) \right] dx^i dx^j \,.
  \ee
Comparing this with the linearly perturbed FRW metric of equation
(\ref{pertFRW}) we can identify the scalar metric perturbations:
  \begin{eqnarray}
   \A=0\\
   B=0 \\
   \psi = \frac{Kr^2}{4}-\frac{\delta a}{a}\\
   E = \frac{K r^4}{16}
  \end{eqnarray}
In addition, the spherical collapse picture is implicitly set up
using comoving spatial coordinates, $x^i$, as the expansion or
collapse is determined solely by the local scale factor,
$\tilde{a}$, and there are no peculiar velocities. Since $B=0$ we
are working in a comoving-orthogonal system with vanishing
3-momentum, $\delta q=0$.

Since $\A=0$, we are also working in a synchronous coordinate system
[which was implicitly assumed in the use of the same time derivative
in Eqs.(\ref{Friedmann}) and~(\ref{pertFriedmann})].
%
%
In a comoving-orthogonal coordinate system the conservation of
momentum requires
 \be
  \A_c = - \frac{\delta P_c}{\rho+P} \,.
 \ee
For pressureless matter momentum conservation thus requires $\A=0$
in a comoving-orthogonal gauge, i.e., the gauge is also synchronous
\cite{1995ApJ...455....7M}, so that the proper cosmic time in the
overdense interior and unperturbed exterior are the same.

Thus we see that the initial data for the spherical collapse problem
which is set up at early times in the matter era where we may use
relativistic perturbation theory is set up in a comoving-orthogonal
gauge. In particular the relevant density perturbation which is used
to determine whether a given region will collapse by the present
time is the comoving-orthogonal density perturbation defined in
Equation~(\ref{deltarhoc}) which is related to the Newtonian
potential by the general relativistic Poisson equation
(\ref{GRPoisson}).

Note that the general relativistic Poisson equation
(\ref{GRPoisson}) shows that for the growing mode solution $\phi_N=
{\rm const.}$ we have $\delta\rho_c\propto a^{-2}$, as the physical
spatial gradients decay in an expanding universe. Thus the density
contrast, $\delta=\delta\rho_c/\rho\propto a$ grows on all scales in
the matter era, as in Newtonian gravity and in the linearised
spherical collapse solution, Equation~(\ref{lineardelta}).


Since the Poisson equation, (\ref{GRPoisson}), relates the Newtonian
potential (\ref{phiN}) to the density perturbation in the
comoving-orthogonal gauge (\ref{deltarhoc}), and the spherical
collapse model uses the density perturbation in the
comoving-orthogonal gauge to determine the collapse of
overdensities, we conclude that the Newtonian form of the
scale-dependent bias does not receive any corrections arising from
general relativity on the super-Hubble scales in the spherical
collapse model.

\subsection{Spherical collapse in longitudinal gauge}

It is of course always possible to consider spherical collapse in
different coordinates.
If we redefine the conformal time and radial coordinate
\begin{eqnarray}
 \eta &=& \left( 1- \frac{1}{20} K(r_0^2-r_\ell^2) \right) \eta_\ell \,, \nonumber\\
 r &=& \left( 1 + \frac{1}{20} K\eta_\ell^2 + K(r_0^2-r_\ell^2) \right) r_\ell
 \,,
 \end{eqnarray}
then the line element (\ref{pertds2}) can be written (to first order
in $K$) in a longitudinal gauge
\begin{equation}
ds^2 = a_\ell^2 \left[ (1-2\phi_N) d\eta_\ell^2 - (1+2\phi_N) \left(
dr_\ell^2 + r_\ell^2 d\Omega^2 \right) \right] \,,
\end{equation}
where $a_\ell=a(\eta_\ell)$ and the Newtonian potential for
$r_\ell<r_0$ is
\begin{equation}
 \phi_N = \frac{3}{20} K (r_0^2-r_\ell^2) \,.
 \end{equation}
Note that we need to specify the size of the overdensity, $r_0$, so
that the coordinate time coincides with the time in the unperturbed
background, $r\geq r_0$.

Note that the density contrast in the longitudinal gauge, given by
Eq.~(\ref{longPoisson}), is no longer spatially homogeneous
 \begin{equation}
 \delta_\ell = \frac35 Ka_\ell \left[ 1+ \frac12 \left(
 \frac{r_0^2-r_\ell^2}{r_H^2} \right) \right] \,,
 \end{equation}
where $r_H={\cal H}^{-1}$ is the comoving Hubble length. The linear
central density contrast at the collapse time is given by
 \begin{equation}
 \label{deltal}
 \delta_{\ell*}|_{r=0} = \frac35 \left( \frac{3\pi}{2} \right)^{2/3} \left[ 1+ \frac12 \left(
 \frac{r_0^2}{r_H^2} \right) \right] \,.
 \end{equation}
In contrast to the result in the comoving-orthogonal gauge,
Eq.~(\ref{deltac}), the criterion for collapse in the longitudinal
gauge is not simply a function of the local density, but it also
depends upon the scale of the overdensity for $r_0\sim r_H$.
Thus, although the GR correction to the Poisson equation in the
longitudinal gauge, Eq.~(\ref{longPoisson}), implies that the
density contrast on super-Hubble scales ($r_0\gg r_H$) becomes a
local function of the Newtonian potential, the collapse condition,
Eq.~(\ref{deltal}), becomes non-local and this is the origin of the
scale-dependent bias as seen in the longitudinal gauge.

More generally, dropping the condition of spherical symmetry, we can
combine the two different expressions of the GR Poisson
equations~(\ref{GRPoisson}) and~(\ref{longPoisson}) to obtain a
manifestly non-local expression for the density contrast in the
longitudinal gauge in terms of the comoving-orthogonal density
contrast
\begin{equation}
\delta_\ell = \left( 1 - 3{\cal H}^2 \partial^{-2} \right) \delta
\,.
\end{equation}

\section{Correlation function}
\label{sec:correlation-function} The necessary consequence of the
fact that there are no general relativistic corrections to the form
of the scale dependent bias on scales larger than the horizon is
that the correlation function formally diverges as we show next.
The correlation function of objects is given by the Fourier
transform of the power spectrum
\begin{equation}
   \xi(r) = \frac{1}{(2\pi)^3}\int \rmd^3k P(k) e^{i\vec{k}\cdot\vec{r}}.
\end{equation}
In the homogeneous and isotropic universe, this can be integrated to
\begin{equation}
\label{eq:2}
   \xi(r) = \frac{1}{2\pi^2} \int P(k) \frac{\sin(kr)}{kr} k^2\rmd k.
\end{equation}

The dark matter power spectrum $P_{\rm dm}(k)=k^{n_s}$ for small
$k$, where $n_s$ is the spectral index of primordial fluctuations.
Therefore, for $P(k)=(b+\Delta b)^2 P_{\rm dm}(k)$, the $P(k)$ is
proportional to $k^{n_s-4}$ (for objects with $b\neq1$) as $k$
approaches zero and hence the integrand is proportional to
$k^{n_s-2}$ in this limit.  The result is that $\xi(r)$ diverges for
\emph{every} $r$.

How does one then relate the measurements of the large scale
structure to the observations in non-Gaussian universes? The hint is
given by the fact, that root mean square of fluctuation in the field
of tracers, i.e. $\sigma = \sqrt{\xi(0)}$ is infinite. However, when
we measure the fluctuations in a given survey, we measure the mean
density from the same survey, effectively forcing the fluctuations
to average to zero. Therefore, the measured density fluctuations
$\tilde{\delta}$ are given by
\begin{equation}
   \tilde{\delta}(\vec{x}) = \delta(\vec{x}) - \bar{\delta},
\end{equation}
where
\begin{equation}
   \bar{\delta} = \int \rmd^3 \vec{x} W(\vec{x}) \delta(\vec{x})
\end{equation}
with $W(\vec{x})$ being the normalised survey window.

This means that the measured correlation function is given by

\begin{multline}
\label{eq:1}
   \tilde{\xi}(r) = \int \rmd^3 \vec{x} W(\vec{x}) \tilde{\delta}(\vec{x})
     \tilde{\delta}(\vec{x}+\vec{r})\\
=\int \rmd^3 \vec{x} W(\vec{x}) \delta(\vec{x})
\delta(\vec{x}+\vec{r}) - \bar{\delta}^2.
\end{multline}

The expectation value for $\tilde{\xi}(r)$, averaged over all
possible realisations of the underlying density fields is thus given
by

\begin{multline}
   \left< \tilde{\xi}(r) \right> =  \int \rmd^3 \vec{x} W(\vec{x})
   \left< \delta(\vec{x}) \delta(\vec{x}+\vec{r})\right> -
   \left< \bar{\delta}^2 \right> \\= \xi(r) - \left< \bar{\delta}^2 \right>.
\end{multline}

The second term of this equation is the variance of $\bar{\delta}$
over an ensemble over possible realisations of the underlying field.
As the universe is ergodic, we can replace many realisations of the
underlying field with many copies of the window function over a
single realisation and hence the second term can be written as
\begin{multline}
\label{eq:3}
   \left<\bar{\delta}^2\right> = \left< (\delta * W)^2(\vec{x}) \right>\\
= \frac{1}{(2 \pi)^3} \int \rmd^3 k W_k(\vec{k})^2 P(k)=\sigma^2_W,
\end{multline}
where $*$ denotes convolution and $W_k$ is the Fourier transform of
the window function. The last term is therefore nothing more than
the variance of the field smoothed over the window function of the
survey. We can therefore write
\begin{equation}
\label{eq:5}
  \left< \tilde{\xi}(r) \right> = \xi(r) - \sigma_W^2.
\end{equation}
This is an expected result. We simply cannot measure variance coming
from scales larger than the survey.

Let us take a concrete example. For the spherical top-hat window
function of size $R$, the integral \eqref{eq:3} becomes
\begin{equation}
   \sigma^2(R) = \frac{1}{2\pi^2}\int \rmd^3 k P(k) k^2 T^2(kR),
\end{equation}
with $T(x) = 3 (\sin x - x \cos x)/x^3$.  In this case, the Equation
(\ref{eq:5}) can be rewritten as
\begin{equation}
  \tilde{\xi}(r) = \frac{1}{2\pi^2} \int P(k) \left(\frac{\sin(kr)}{kr}-T^2(kR)\right) k^2\rmd k.
\end{equation}
This is now a well behaved transform. In the limit of $k\rightarrow
0$, the term in brackets becomes $k^2(R^2/10-r^2/6)$ and hence the
integrand is proportional to $k^{n_s}$ and thus convergent.

Note that this cancellation occurs for any finite window function
$W$. If a window function has a largest typical linear scale $R$, then for
$k \ll R^{-1}$, $W_k\sim1$ as the $\exp(ikR)$ is varying slowly
across the window. One these scales $\sin(kr)/{kr}-W^2(kR)$ is
therefore quadratic in $k$ to the leading order.

Finally, there are two other worries associated with a diverging
field. First, one might worry that the are formal problems with
perturbing a homogeneous universe with a solution that has infra-red
divergence. However, the underlying physical field, the dark-matter
density, is a completely regular and does not diverge on the large
scales. Only some non-linear transformation of this regular field,
namely a biased tracer of this field, or a field of peak positions
(after smoothing on some scale) has a large-scale divergence.
Therefore, this should not be a problem. A more serious issue is
that the field of tracers, although it can be arbitrarily biased
with respect to the underlying dark matter density field, must still
remain positive, i.e. $\delta(\vec{x})>-1$ since the local density
of astrophysical objects cannot be negative. To estimate when this
becomes important, we demand that the dimensionless power per unit
log $k$, $\Delta^2 = P(k)k^3/4\pi^2$ is much less than unity on
large scales. For biased tracer, this reduces a requirement that
$f_{\rm NL}(b-1)\ll10^4$, which is always true for observationally
interesting values of $\fnl$ and $b$.

\section{Conclusions}
\label{sec:conclusions}

In this paper we have discussed the derivation of scale-dependent
bias from primordial non-Gaussianity in the context of general
relativity. Our results are somewhat unexciting - we find no
corrections arising from general relativistic effects on scales
larger than the Hubble length. We have justified this by noting that
the spherical collapse and thus peak-background split formalisms
that are performed in the Newtonian context can be carried over into
general relativity if one works in terms of the density perturbation
in the comoving-orthogonal gauge. This justifies the use of the
Newtonian form of the Poisson equation (\ref{Poisson}) on all scales
(in contrast to what one might naively think
\cite{2008arXiv0808.2689D}).

This means that the power spectrum of biased tracers of the density
field diverges as the scales of interest become larger and larger
and leads to the infinite variance in the local density of the
biased objects. The formal correlation function is divergent on all
scales, but any observed correlation function will be finite,
because variance on scales larger than the size of the survey will
not be observed.

The reason that general relativity reduces to apparently Newtonian
equations is because we are only considering scalar perturbations
which obey energy and momentum constraint equations relating the
metric to the matter perturbations. Considering first-order
perturbations about an FRW background, the constraint equations
enforce a Poisson-type relation between the local density
perturbation and the metric perturbations on a given constant-time
hypersurface, coinciding with a given time in the background
spacetime. Coordinate freedom in general relativity allows different
choices of time slicing in the perturbed spacetime, but it is the
density perturbation in the comoving-orthogonal gauge that follows
the same behaviour as the Newtonian density perturbation on all
scales.

Assuming spherical symmetry we are able to follow the non-linear
collapse in a comoving-orthogonal coordinate system. Spherical
symmetry eliminates the vector and tensor parts of the metric
perturbation which might have introduced non-Newtonian behaviour.
For example, there are no gravitational waves if we impose spherical
symmetry. Deviations from spherical symmetry could thus introduce
additional general relativistic terms, but the scalar, vector and
tensor perturbations are decoupled from scalar density perturbations
at first-order, so such effects would only be expected to arise at
second- or higher-order \cite{2007PhRvD..76j3527H}. This might
ultimately be important when we wish to constrain non-linearity
parameters, $\fNL$, of order unity but it suggests that general
relativistic corrections to the Newtonian results currently being
used are negligible for current data.

\section*{Acknowledgements}
Authors thank the Galileo Galilei Institute for Theoretical Physics
for their hospitality and the INFN for partial support during the
completion of this work. DW is grateful to Rob Crittenden, Kazuya
Koyama, David Lyth and Will Percival for helpful discussions. AS
acknowledges useful discussions with Martin White. DW is supported by
the STFC, while AS is supported by Berkeley Center for Cosmological
Physics.

\bibliographystyle{arxiv}
\bibliography{cosmo,cosmo_preprints,cosmo_wands}

\def\eprinttmppp@#1arXiv:@{#1}
\providecommand{\arxivlink[1]}{\href{http://arxiv.org/abs/#1}{arXiv:#1}}
\def\eprinttmp@#1arXiv:#2 [#3]#4@{\ifthenelse{\equal{#3}{x}}{\ifthenelse{
\equal{#1}{}}{\arxivlink{\eprinttmppp@#2@}}{\arxivlink{#1}}}{\arxivlink{#2}
  [#3]}}
\providecommand{\eprintlink}[1]{\eprinttmp@#1arXiv: [x]@}
\renewcommand{\eprint}[1]{\eprintlink{#1}}
\providecommand{\adsurl}[1]{\href{#1}{ADS}}
\renewcommand{\bibinfo}[2]{\ifthenelse{\equal{#1}{isbn}}{\href{http://cosmolog%
ist.info/ISBN/#2}{#2}}{#2}}
\begin{thebibliography}{29}
\expandafter\ifx\csname natexlab\endcsname\relax\def\natexlab#1{#1}\fi
\expandafter\ifx\csname bibnamefont\endcsname\relax
  \def\bibnamefont#1{#1}\fi
\expandafter\ifx\csname bibfnamefont\endcsname\relax
  \def\bibfnamefont#1{#1}\fi
\expandafter\ifx\csname citenamefont\endcsname\relax
  \def\citenamefont#1{#1}\fi
\expandafter\ifx\csname url\endcsname\relax
  \def\url#1{\texttt{#1}}\fi
\expandafter\ifx\csname urlprefix\endcsname\relax\def\urlprefix{URL }\fi

\bibitem[{\citenamefont{{Salopek} and {Bond}}(1990)}]{1990PhRvD..42.3936S}
\bibinfo{author}{\bibfnamefont{D.~S.} \bibnamefont{{Salopek}}}
  \bibnamefont{and} \bibinfo{author}{\bibfnamefont{J.~R.}
  \bibnamefont{{Bond}}}, \bibinfo{journal}{\prd} \textbf{\bibinfo{volume}{42}},
  \bibinfo{pages}{3936} (\bibinfo{year}{1990}),
  \adsurl{http://adsabs.harvard.edu/abs/1990PhRvD..42.3936S}.

\bibitem[{\citenamefont{{Gangui} et~al.}(1994)\citenamefont{{Gangui},
  {Lucchin}, {Matarrese}, and {Mollerach}}}]{1994ApJ...430..447G}
\bibinfo{author}{\bibfnamefont{A.}~\bibnamefont{{Gangui}}},
  \bibinfo{author}{\bibfnamefont{F.}~\bibnamefont{{Lucchin}}},
  \bibinfo{author}{\bibfnamefont{S.}~\bibnamefont{{Matarrese}}},
  \bibnamefont{and}
  \bibinfo{author}{\bibfnamefont{S.}~\bibnamefont{{Mollerach}}},
  \bibinfo{journal}{\apj} \textbf{\bibinfo{volume}{430}}, \bibinfo{pages}{447}
  (\bibinfo{year}{1994}), \eprint{arXiv:astro-ph/9312033}.

\bibitem[{\citenamefont{{Verde} et~al.}(2000)\citenamefont{{Verde}, {Wang},
  {Heavens}, and {Kamionkowski}}}]{2000MNRAS.313..141V}
\bibinfo{author}{\bibfnamefont{L.}~\bibnamefont{{Verde}}},
  \bibinfo{author}{\bibfnamefont{L.}~\bibnamefont{{Wang}}},
  \bibinfo{author}{\bibfnamefont{A.~F.} \bibnamefont{{Heavens}}},
  \bibnamefont{and}
  \bibinfo{author}{\bibfnamefont{M.}~\bibnamefont{{Kamionkowski}}},
  \bibinfo{journal}{\mnras} \textbf{\bibinfo{volume}{313}},
  \bibinfo{pages}{141} (\bibinfo{year}{2000}), \eprint{arXiv:astro-ph/9906301}.

\bibitem[{\citenamefont{{Lyth} and
  {Rodr{\'{\i}}guez}}(2005)}]{2005PhRvL..95l1302L}
\bibinfo{author}{\bibfnamefont{D.~H.} \bibnamefont{{Lyth}}} \bibnamefont{and}
  \bibinfo{author}{\bibfnamefont{Y.}~\bibnamefont{{Rodr{\'{\i}}guez}}},
  \bibinfo{journal}{Phys. Rev. Lett.} \textbf{\bibinfo{volume}{95}},
  \bibinfo{pages}{121302} (\bibinfo{year}{2005}),
  \eprint{arXiv:astro-ph/0504045}.

\bibitem[{\citenamefont{{Lyth} et~al.}(2003)\citenamefont{{Lyth}, {Ungarelli},
  and {Wands}}}]{2003PhRvD..67b3503L}
\bibinfo{author}{\bibfnamefont{D.~H.} \bibnamefont{{Lyth}}},
  \bibinfo{author}{\bibfnamefont{C.}~\bibnamefont{{Ungarelli}}},
  \bibnamefont{and} \bibinfo{author}{\bibfnamefont{D.}~\bibnamefont{{Wands}}},
  \bibinfo{journal}{\prd} \textbf{\bibinfo{volume}{67}},
  \bibinfo{pages}{023503} (\bibinfo{year}{2003}),
  \eprint{arXiv:astro-ph/0208055}.

\bibitem[{\citenamefont{{Creminelli} and
  {Senatore}}(2007)}]{2007JCAP...11..010C}
\bibinfo{author}{\bibfnamefont{P.}~\bibnamefont{{Creminelli}}}
  \bibnamefont{and}
  \bibinfo{author}{\bibfnamefont{L.}~\bibnamefont{{Senatore}}},
  \bibinfo{journal}{Journal of Cosmology and Astro-Particle Physics}
  \textbf{\bibinfo{volume}{11}}, \bibinfo{pages}{10} (\bibinfo{year}{2007}),
  \eprint{arXiv:hep-th/0702165}.

\bibitem[{\citenamefont{{Koyama} et~al.}(2007)\citenamefont{{Koyama}, {Mizuno},
  {Vernizzi}, and {Wands}}}]{2007JCAP...11..024K}
\bibinfo{author}{\bibfnamefont{K.}~\bibnamefont{{Koyama}}},
  \bibinfo{author}{\bibfnamefont{S.}~\bibnamefont{{Mizuno}}},
  \bibinfo{author}{\bibfnamefont{F.}~\bibnamefont{{Vernizzi}}},
  \bibnamefont{and} \bibinfo{author}{\bibfnamefont{D.}~\bibnamefont{{Wands}}},
  \bibinfo{journal}{Journal of Cosmology and Astro-Particle Physics}
  \textbf{\bibinfo{volume}{11}}, \bibinfo{pages}{24} (\bibinfo{year}{2007}),
  \eprint{0708.4321}.

\bibitem[{\citenamefont{{Buchbinder} et~al.}(2008)\citenamefont{{Buchbinder},
  {Khoury}, and {Ovrut}}}]{2008PhRvL.100q1302B}
\bibinfo{author}{\bibfnamefont{E.~I.} \bibnamefont{{Buchbinder}}},
  \bibinfo{author}{\bibfnamefont{J.}~\bibnamefont{{Khoury}}}, \bibnamefont{and}
  \bibinfo{author}{\bibfnamefont{B.~A.} \bibnamefont{{Ovrut}}},
  \bibinfo{journal}{Phys. Rev. Lett.} \textbf{\bibinfo{volume}{100}},
  \bibinfo{pages}{171302} (\bibinfo{year}{2008}), \eprint{arXiv:0710.5172}.

\bibitem[{\citenamefont{{Lehners} and
  {Steinhardt}}(2008)}]{2008PhRvD..77f3533L}
\bibinfo{author}{\bibfnamefont{J.-L.} \bibnamefont{{Lehners}}}
  \bibnamefont{and} \bibinfo{author}{\bibfnamefont{P.~J.}
  \bibnamefont{{Steinhardt}}}, \bibinfo{journal}{\prd}
  \textbf{\bibinfo{volume}{77}}, \bibinfo{pages}{063533}
  (\bibinfo{year}{2008}), \eprint{arXiv:0712.3779}.

\bibitem[{\citenamefont{{Komatsu} and {Spergel}}(2001)}]{2001PhRvD..63f3002K}
\bibinfo{author}{\bibfnamefont{E.}~\bibnamefont{{Komatsu}}} \bibnamefont{and}
  \bibinfo{author}{\bibfnamefont{D.~N.} \bibnamefont{{Spergel}}},
  \bibinfo{journal}{\prd} \textbf{\bibinfo{volume}{63}},
  \bibinfo{pages}{063002} (\bibinfo{year}{2001}),
  \eprint{arXiv:astro-ph/0005036}.

\bibitem[{\citenamefont{{Maldacena}}(2003)}]{2003JHEP...05..013M}
\bibinfo{author}{\bibfnamefont{J.}~\bibnamefont{{Maldacena}}},
  \bibinfo{journal}{Journal of High Energy Physics}
  \textbf{\bibinfo{volume}{5}}, \bibinfo{pages}{13} (\bibinfo{year}{2003}),
  \eprint{arXiv:astro-ph/0210603}.

\bibitem[{\citenamefont{{Mukhanov} et~al.}(1992)\citenamefont{{Mukhanov},
  {Feldman}, and {Brandenberger}}}]{1992PhR...215..203M}
\bibinfo{author}{\bibfnamefont{V.~F.} \bibnamefont{{Mukhanov}}},
  \bibinfo{author}{\bibfnamefont{H.~A.} \bibnamefont{{Feldman}}},
  \bibnamefont{and} \bibinfo{author}{\bibfnamefont{R.~H.}
  \bibnamefont{{Brandenberger}}}, \bibinfo{journal}{\physrep}
  \textbf{\bibinfo{volume}{215}}, \bibinfo{pages}{203} (\bibinfo{year}{1992}),
  \adsurl{http://adsabs.harvard.edu/abs/1992PhR...215..203M}.

\bibitem[{\citenamefont{{Press} and {Schechter}}(1974)}]{1974ApJ...187..425P}
\bibinfo{author}{\bibfnamefont{W.~H.} \bibnamefont{{Press}}} \bibnamefont{and}
  \bibinfo{author}{\bibfnamefont{P.}~\bibnamefont{{Schechter}}},
  \bibinfo{journal}{\apj} \textbf{\bibinfo{volume}{187}}, \bibinfo{pages}{425}
  (\bibinfo{year}{1974}),
  \adsurl{http://adsabs.harvard.edu/abs/1974ApJ...187..425P}.

\bibitem[{\citenamefont{{Efstathiou} et~al.}(1988)\citenamefont{{Efstathiou},
  {Frenk}, {White}, and {Davis}}}]{1988MNRAS.235..715E}
\bibinfo{author}{\bibfnamefont{G.}~\bibnamefont{{Efstathiou}}},
  \bibinfo{author}{\bibfnamefont{C.~S.} \bibnamefont{{Frenk}}},
  \bibinfo{author}{\bibfnamefont{S.~D.~M.} \bibnamefont{{White}}},
  \bibnamefont{and} \bibinfo{author}{\bibfnamefont{M.}~\bibnamefont{{Davis}}},
  \bibinfo{journal}{\mnras} \textbf{\bibinfo{volume}{235}},
  \bibinfo{pages}{715} (\bibinfo{year}{1988}),
  \adsurl{http://adsabs.harvard.edu/abs/1988MNRAS.235..715E}.

\bibitem[{\citenamefont{{Cole} and {Kaiser}}(1989)}]{1989MNRAS.237.1127C}
\bibinfo{author}{\bibfnamefont{S.}~\bibnamefont{{Cole}}} \bibnamefont{and}
  \bibinfo{author}{\bibfnamefont{N.}~\bibnamefont{{Kaiser}}},
  \bibinfo{journal}{\mnras} \textbf{\bibinfo{volume}{237}},
  \bibinfo{pages}{1127} (\bibinfo{year}{1989}),
  \adsurl{http://adsabs.harvard.edu/abs/1989MNRAS.237.1127C}.

\bibitem[{\citenamefont{{Mo} and {White}}(1996)}]{1996MNRAS.282..347M}
\bibinfo{author}{\bibfnamefont{H.~J.} \bibnamefont{{Mo}}} \bibnamefont{and}
  \bibinfo{author}{\bibfnamefont{S.~D.~M.} \bibnamefont{{White}}},
  \bibinfo{journal}{\mnras} \textbf{\bibinfo{volume}{282}},
  \bibinfo{pages}{347} (\bibinfo{year}{1996}),
  \adsurl{http://adsabs.harvard.edu/cgi-bin/nph-bib_query?bibcode=1996MNRAS.28%
2..347M&amp;db_key=AST}.

\bibitem[{\citenamefont{{Catelan} et~al.}(1998)\citenamefont{{Catelan},
  {Lucchin}, {Matarrese}, and {Porciani}}}]{1998MNRAS.297..692C}
\bibinfo{author}{\bibfnamefont{P.}~\bibnamefont{{Catelan}}},
  \bibinfo{author}{\bibfnamefont{F.}~\bibnamefont{{Lucchin}}},
  \bibinfo{author}{\bibfnamefont{S.}~\bibnamefont{{Matarrese}}},
  \bibnamefont{and}
  \bibinfo{author}{\bibfnamefont{C.}~\bibnamefont{{Porciani}}},
  \bibinfo{journal}{\mnras} \textbf{\bibinfo{volume}{297}},
  \bibinfo{pages}{692} (\bibinfo{year}{1998}), \eprint{arXiv:astro-ph/9708067}.

\bibitem[{\citenamefont{{Dalal} et~al.}(2008)\citenamefont{{Dalal}, {Dor{\'e}},
  {Huterer}, and {Shirokov}}}]{Dalal:2007cu}
\bibinfo{author}{\bibfnamefont{N.}~\bibnamefont{{Dalal}}},
  \bibinfo{author}{\bibfnamefont{O.}~\bibnamefont{{Dor{\'e}}}},
  \bibinfo{author}{\bibfnamefont{D.}~\bibnamefont{{Huterer}}},
  \bibnamefont{and}
  \bibinfo{author}{\bibfnamefont{A.}~\bibnamefont{{Shirokov}}},
  \bibinfo{journal}{\prd} \textbf{\bibinfo{volume}{77}},
  \bibinfo{pages}{123514} (\bibinfo{year}{2008}), \eprint{arXiv:0710.4560}.

\bibitem[{\citenamefont{{Slosar}
  et~al.}(2008{\natexlab{a}})\citenamefont{{Slosar}, {Hirata}, {Seljak}, {Ho},
  and {Padmanabhan}}}]{2008JCAP...08..031S}
\bibinfo{author}{\bibfnamefont{A.}~\bibnamefont{{Slosar}}},
  \bibinfo{author}{\bibfnamefont{C.}~\bibnamefont{{Hirata}}},
  \bibinfo{author}{\bibfnamefont{U.}~\bibnamefont{{Seljak}}},
  \bibinfo{author}{\bibfnamefont{S.}~\bibnamefont{{Ho}}}, \bibnamefont{and}
  \bibinfo{author}{\bibfnamefont{N.}~\bibnamefont{{Padmanabhan}}},
  \bibinfo{journal}{Journal of Cosmology and Astro-Particle Physics}
  \textbf{\bibinfo{volume}{8}}, \bibinfo{pages}{31}
  (\bibinfo{year}{2008}{\natexlab{a}}), \eprint{0805.3580}.

\bibitem[{\citenamefont{{Matarrese} and {Verde}}(2008)}]{2008ApJ...677L..77M}
\bibinfo{author}{\bibfnamefont{S.}~\bibnamefont{{Matarrese}}} \bibnamefont{and}
  \bibinfo{author}{\bibfnamefont{L.}~\bibnamefont{{Verde}}},
  \bibinfo{journal}{\apjl} \textbf{\bibinfo{volume}{677}}, \bibinfo{pages}{L77}
  (\bibinfo{year}{2008}), \eprint{arXiv:0801.4826}.

\bibitem[{\citenamefont{{Slosar}
  et~al.}(2008{\natexlab{b}})\citenamefont{{Slosar}, {Hirata}, {Seljak}, {Ho},
  and {Padmanabhan}}}]{2008arXiv0805.3580S}
\bibinfo{author}{\bibfnamefont{A.}~\bibnamefont{{Slosar}}},
  \bibinfo{author}{\bibfnamefont{C.}~\bibnamefont{{Hirata}}},
  \bibinfo{author}{\bibfnamefont{U.}~\bibnamefont{{Seljak}}},
  \bibinfo{author}{\bibfnamefont{S.}~\bibnamefont{{Ho}}}, \bibnamefont{and}
  \bibinfo{author}{\bibfnamefont{N.}~\bibnamefont{{Padmanabhan}}},
  \bibinfo{journal}{ArXiv e-prints} \textbf{\bibinfo{volume}{805}}
  (\bibinfo{year}{2008}{\natexlab{b}}), \eprint{0805.3580}.

\bibitem[{\citenamefont{{Afshordi} and {Tolley}}(2008)}]{2008arXiv0806.1046A}
\bibinfo{author}{\bibfnamefont{N.}~\bibnamefont{{Afshordi}}} \bibnamefont{and}
  \bibinfo{author}{\bibfnamefont{A.~J.} \bibnamefont{{Tolley}}},
  \bibinfo{journal}{ArXiv e-prints} \textbf{\bibinfo{volume}{806}}
  (\bibinfo{year}{2008}), \eprint{0806.1046}.

\bibitem[{\citenamefont{{McDonald}}(2008)}]{2008arXiv0806.1061M}
\bibinfo{author}{\bibfnamefont{P.}~\bibnamefont{{McDonald}}},
  \bibinfo{journal}{ArXiv e-prints} \textbf{\bibinfo{volume}{806}}
  (\bibinfo{year}{2008}), \eprint{0806.1061}.

\bibitem[{\citenamefont{{Malik} and {Wands}}(2008)}]{2008arXiv0809.4944M}
\bibinfo{author}{\bibfnamefont{K.~A.} \bibnamefont{{Malik}}} \bibnamefont{and}
  \bibinfo{author}{\bibfnamefont{D.}~\bibnamefont{{Wands}}},
  \bibinfo{journal}{ArXiv e-prints}  (\bibinfo{year}{2008}),
  \eprint{0809.4944}.

\bibitem[{\citenamefont{{Hu} et~al.}(1998)\citenamefont{{Hu}, {Seljak},
  {White}, and {Zaldarriaga}}}]{1998PhRvD..57.3290H}
\bibinfo{author}{\bibfnamefont{W.}~\bibnamefont{{Hu}}},
  \bibinfo{author}{\bibfnamefont{U.}~\bibnamefont{{Seljak}}},
  \bibinfo{author}{\bibfnamefont{M.}~\bibnamefont{{White}}}, \bibnamefont{and}
  \bibinfo{author}{\bibfnamefont{M.}~\bibnamefont{{Zaldarriaga}}},
  \bibinfo{journal}{\prd} \textbf{\bibinfo{volume}{57}}, \bibinfo{pages}{3290}
  (\bibinfo{year}{1998}),
  \adsurl{http://adsabs.harvard.edu/cgi-bin/nph-bib_query?bibcode=1998PhRvD..5%
7.3290H&db_k ey=AST}.

\bibitem[{\citenamefont{{Bardeen}}(1980)}]{1980PhRvD..22.1882B}
\bibinfo{author}{\bibfnamefont{J.~M.} \bibnamefont{{Bardeen}}},
  \bibinfo{journal}{\prd} \textbf{\bibinfo{volume}{22}}, \bibinfo{pages}{1882}
  (\bibinfo{year}{1980}),
  \adsurl{http://adsabs.harvard.edu/cgi-bin/nph-bib_query?bibcode=1980PhRvD..2%
2.1882B&db_key=AST}.

\bibitem[{\citenamefont{{Ma} and {Bertschinger}}(1995)}]{1995ApJ...455....7M}
\bibinfo{author}{\bibfnamefont{C.}~\bibnamefont{{Ma}}} \bibnamefont{and}
  \bibinfo{author}{\bibfnamefont{E.}~\bibnamefont{{Bertschinger}}},
  \bibinfo{journal}{\apj} \textbf{\bibinfo{volume}{455}}, \bibinfo{pages}{7}
  (\bibinfo{year}{1995}),
  \adsurl{http://adsabs.harvard.edu/cgi-bin/nph-bib_query?bibcode=1995ApJ...45%
5....7M&amp;db_key=AST}.

\bibitem[{\citenamefont{{Dent} and {Dutta}}(2008)}]{2008arXiv0808.2689D}
\bibinfo{author}{\bibfnamefont{J.~B.} \bibnamefont{{Dent}}} \bibnamefont{and}
  \bibinfo{author}{\bibfnamefont{S.}~\bibnamefont{{Dutta}}},
  \bibinfo{journal}{ArXiv e-prints}  (\bibinfo{year}{2008}),
  \eprint{0808.2689}.

\bibitem[{\citenamefont{{Hwang} and {Noh}}(2007)}]{2007PhRvD..76j3527H}
\bibinfo{author}{\bibfnamefont{J.-C.} \bibnamefont{{Hwang}}} \bibnamefont{and}
  \bibinfo{author}{\bibfnamefont{H.}~\bibnamefont{{Noh}}},
  \bibinfo{journal}{\prd} \textbf{\bibinfo{volume}{76}},
  \bibinfo{pages}{103527} (\bibinfo{year}{2007}), \eprint{0704.1927}.

\end{thebibliography}

\end{document}